\documentclass[fleqn,usenatbib]{mnras}
\usepackage{graphicx}
\usepackage{dcolumn}
\usepackage{bm}
\usepackage{natbib}
\usepackage{amsmath, amssymb}
\usepackage[utf8]{inputenc}
\usepackage{hyperref}

\makeatother

\title[The chaotic four-body problem]{The chaotic four-body problem in Newtonian gravity - II. An ansatz-based approach to analytic solutions}
\author[Carlos M. Barrera Retamal, Nathan W.C. Leigh, Nicholas C. Stone]{Carlos M. Barrera Retamal$^{1}$, Nathan W.C. Leigh$^{1,2}$, Nicholas C. Stone$^3$
\\
$^{1}$Departamento de Astronom\'ia, Facultad Ciencias F\'isicas y Matem\'aticas, Universidad de Concepci\'on, Chile\\
$^2$Department of Astrophysics, American Museum of Natural History, Central Park West and 79th Street, New York, NY, 10024, USA\\
$^3$Columbia Astrophysics Laboratory, Columbia University, New York, NY, 10027, USA
}

\date{Accepted XXX. Received YYY; in original form ZZZ}

\pubyear{2022}

\begin{document}
\label{firstpage}
\pagerange{\pageref{firstpage}--\pageref{lastpage}}
\maketitle

\begin{abstract}
In this paper, we continue our analysis of the chaotic four-body problem by presenting a general ansatz-based analytic treatment using statistical mechanics, where each outcome of the four-body problem is regarded as some variation of the three-body problem (e.g., when two single stars are produced, called the 2+1+1 outcome, each ejection event is modeled as its own three-body interaction by assuming that the ejections are well separated in time). This is a generalization of the statistical mechanics treatment of the three-body problem based on the density-of-states formalism. In our case, we focus on the interaction of two binary systems, after which we divide our results into three possible outcome scenarios (2+2, 2+1+1, and 3+1). For each outcome, we apply an ansatz-based approach to deriving analytic distribution functions that describe the properties of the products of chaotic four-body interactions involving point particles.  To test our theoretical distributions, we perform a set of scattering simulations in the equal-mass point particle limit using \texttt{FEWBODY}. We compare our final theoretical distributions to the simulations for each particular scenario, finding consistently good agreement between the two.  The highlights of our results include that binary-binary scatterings act to systematically destroy binaries producing instead a single binary and two ejected stars or a stable triple, the 2+2 outcome produces the widest binaries and the 2+1+1 outcome produces the most compact binaries.  

\end{abstract}

\date{\today}
\maketitle

\section{Introduction}

The four-body problem has scarcely been studied analytically.  This can be understood, at least in part, upon considering the history of the three-body problem, and its notoriety for being a strong example of chaos in nature.  With the two-body problem solved, the temptation to find an analytic solution to the three-body problem attracted many researchers.  Generally, the goal was to predict the positions of the particles at any future time, for any set of initial conditions.  This eventually led to the understanding that the addition of even one extra particle (relative to the two-body problem) renders the number of variables in the equations of motion greater than the number of equations.  The problem is unsolvable, and more particles will only make it worse.  Consequently, the four-body problem received little attention for centuries \citep[e.g.][]{nash80}.

More recently, computational advances have allowed for numerical studies of the four-body problem \citep[e.g.][]{harrington74,saslaw74,mikkola83,mikkola84a,mikkola84b,rasio95,fregeau04,leigh12,Leigh+16,ryu17a,ryu17b,ryu17c}.  Ignoring planetary dynamics, most of these studied scattering interactions between two binary star systems.  For example, \citet{mikkola83} confirmed that stable triple systems form during encounters between identical binaries.  \citet{mikkola84a} extended this result to include binaries with different initial orbital energies, finding in the process that significantly more triples form as the ratio of binding energies increases from unity.

The primary astrophysical motivation for this paper is binary-binary scatterings in dense stellar systems, such as open, globular, or nuclear star clusters.  In such systems, the rate of binary-binary scatterings can dominate the rate of binary-single scatterings provided the binary fraction satisfies f$_{\rm b} \gtrsim$ 10\% \citep{sigurdsson93,leigh11}.  In this case, binary-binary scatterings are the dominant cluster heating source and the 4-body problem's scattering outcomes become critical for understanding the thermodynamic evolution of the host star cluster.  Moreover, one possible outcome of a binary-binary scattering (which does not occur for binary-single scatterings in the point particle limit) is the dynamical formation of a stable triple star system \citep[e.g.][]{Leigh+16}.  Such triple systems are of great interest for their susceptibility to the Kozai-Lidov mechanism, which can create accreting inner binaries or exotic astrophysical transients \citep[e.g.][]{perets09}.  The decay products of binary-binary scatterings are thought to have been observed directly, both in the form of stable triples \citep[e.g.][]{leigh11,leigh13} and runaway O/B stars \citep[e.g.][]{hoogerwerf01,oh15}.  

The strongly chaotic nature of the generic four-body problem makes an analytic solution to a set of specific initial conditions impossible (except for some fine-tuned sets of measure zero).  However, chaos becomes a useful tool if we are interested in {\it probabilistic distributions} of outcomes corresponding to {\it distributions} of initial conditions.  Following the pioneering work of \citet[hereafter the ``Monaghan formalism'']{Monaghan76a, Monaghan76b}, we employ statistical mechanics to compute distributions of outcomes for the generic problem of binary-binary scatterings.  This paper is a more analytic continuation of our previous work, which examined binary-binary scattering using a large suite of numerical integrations \citep[hereafter ``Paper I'']{Leigh+16}.

{In \S \ref{sec:setup}, we provide an overview of the geometry and phase space of the problem.  In {\S} \ref{sec:2p2}, \ref{sec:3p1}, and \ref{sec:2p1p1}, we use the Monaghan formalism to compute phase space volumes and parameter distributions for the three possible outcomes of a binary-binary scattering event involving point particles.  In {\S} \ref{sec:branching}, we compute branching ratios for these three outcomes. In {\S} \ref{sec:methods} we present the toolkit \texttt{FEWBODY} to describe and explain the numerical simulations we use to test our analytic distributions. In {\S} \ref{sec:results} we show  our results comparing the simulated data with the analytic derivations found in the previous sections. In {\S} \ref{sec:discuss} we consider the astrophysical implications of our results, and in \S \ref{sec:conclusions} we summarize our work.}

\section{Statistical Mechanics of the Four-Body Problem}
\label{sec:setup}

In this paper we consider the outcome of the generic, non-hierarchical four-body problem.  We assume all stars are point particles and neglect tidal forces, physical collisions, the effects of general relativity, and non-gravitational forces.  The four interacting stars have masses $m_{\rm a}$, $m_{\rm b}$, $m_{\rm c}$, and $m_{\rm d}$, which may differ from each other.  The interacting system has a conserved total energy, $E_0$, and a conserved total angular momentum, $\vec{L}_0$.  Our approach, which was first developed in the Monaghan formalism to treat the chaotic three-body problem, is to consider the statistical phase space of different outcomes of the four-body problem.  By calculating the phase space volume of a single outcome as a function of parameters of interest (e.g. ejection velocities), we can construct distributions of these outcome parameters.  By calculating the relative volumes of different outcomes, we can compute branching ratios between them.

There are four possible outcomes of the generic four-body problem.  Following a phase of chaotic, non-hierarchical gravitational interactions, the system eventually forms some combination of hierarchical, bound particles and unbound, escaping particles.  The four possible combinations are 
\begin{enumerate}
\item One escaping, unbound star and a hierarchically stable triple (the ``3+1'' outcome).
\item Two escaping, unbound stars and a surviving binary (the ``2+1+1'' outcome).
\item Two surviving binaries which are mutually unbound from each other (the ``2+2'' outcome). 
\item Four escaping, unbound stars (the ``1+1+1+1'' outcome).
\end{enumerate}
In isothermal star clusters with Maxwellian velocity distributions, the 1+1+1+1 outcome is almost always energetically forbidden \citep{Leigh+16}, so we ignore it for the remainder of this work.  

The primary assumption we make in analyzing the 3+1, 2+1+1, and 2+2 cases is that the strongly chaotic interactions of a non-hierarchical four-body system will uniformly populate the phase space available in each of these outcomes.  The same assumption motivated the statistical mechanical treatment of the three-body problem in \citet{Monaghan76a, Monaghan76b}, and was validated by post-hoc checks using numerical orbit integrations.  Because the parameter space of the four-body problem is much larger than that of the three-body problem, we cannot hope to fully cover it with numerical integrations, but we will check our results against a suite of 
four-body scattering experiments.

Although our primary motivation is to make statistical predictions for the outcomes of binary-binary scatterings, our results should apply equally well to other non-hierarchical four-body systems, such as a strong triple-single scattering, or to other, less probable events (such as a simultaneous encounter between a binary and two single stars).  

In the following three sections, we consider the 2+2, 3+1, and 2+1+1 outcomes.  It is important to note that variables in these sections may sometimes have the same names but different definitions.  We define all variables locally at the beginning of their respective sections.

\section{The 2+2 Outcome}
\label{sec:2p2}
\begin{figure}
\includegraphics[width=85mm]{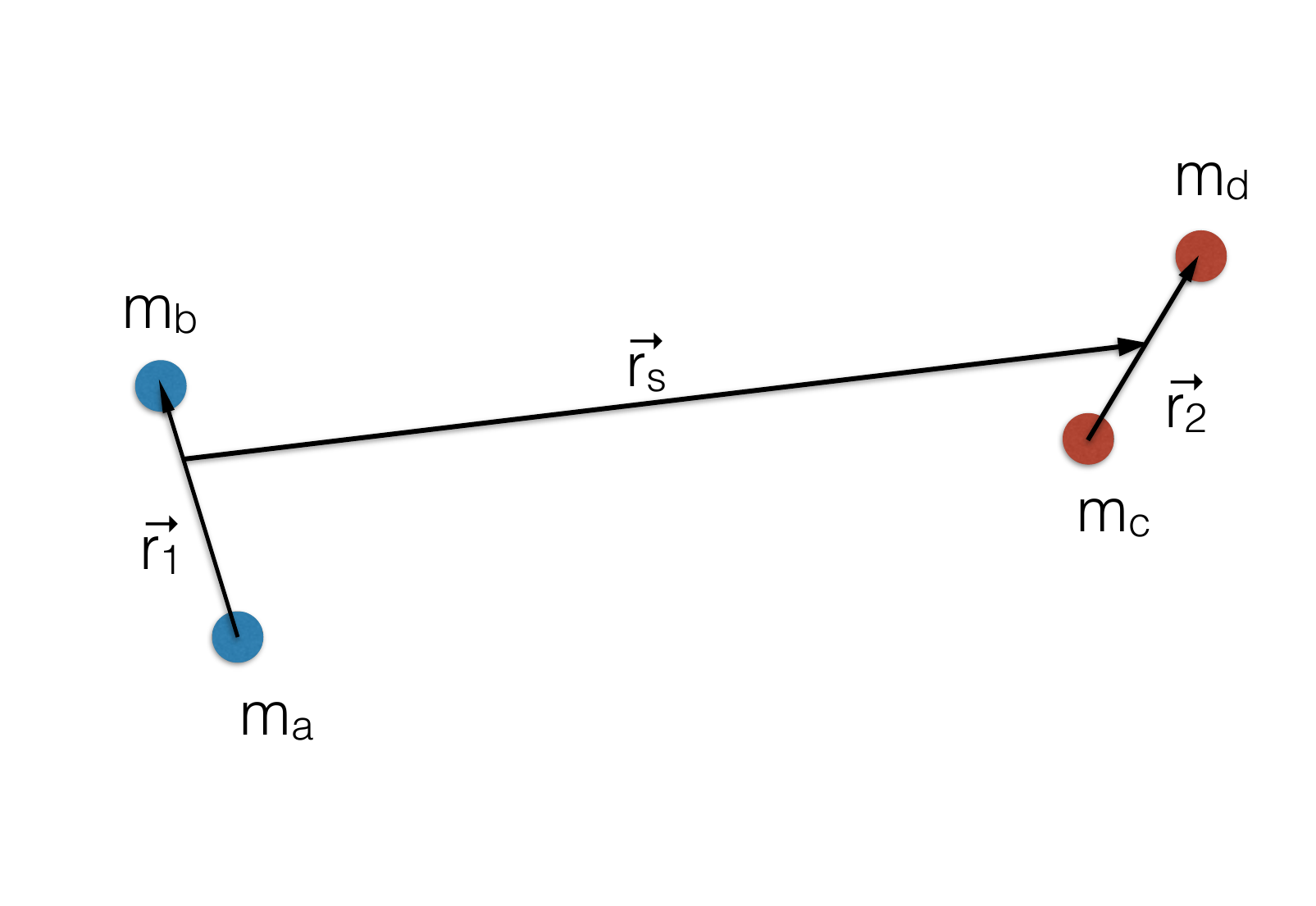}
\caption{The configuration of the 2+2 outcome.}
\label{fig:2p2}
\end{figure}

We begin with the 2+2 case, as it has the greatest similarity to the classic 2+1 three-body problem.  Fig. \ref{fig:2p2} shows a cartoon of the 2+2 outcome.  The final state here is two binaries; the first, composed of masses $m_{\rm a}$ and $m_{\rm b}$, has total mass $m_{\rm B1}=m_{\rm a}+m_{\rm b}$.  The second, composed of masses $m_{\rm c}$ and $m_{\rm d}$, has total mass $m_{\rm B2}=m_{\rm c}+m_{\rm d}$.  The binaries have individual reduced masses $\mathcal{M}_1=m_{\rm a} m_{\rm b}/m_{\rm B1}$ and $\mathcal{M}_2=m_{\rm c} m_{\rm d}/m_{\rm B2}$, and a joint reduced mass of $m=m_{\rm B1}m_{\rm B2}/M$, where $M=m_{\rm B1}+m_{\rm B2}$.

The total energy of this final state is $E_0 = E_{\rm s}+E_{\rm B1}+E_{\rm B2}$, where
\begin{align}
E_{\rm s} =& \frac{1}{2}m\dot{\vec{r}}_{\rm s}^{~2}-\frac{Gm_{\rm B1}m_{\rm B2}}{r_{\rm s}} \notag \\
E_{\rm B1} =& \frac{1}{2}\mathcal{M}_1\dot{\vec{r}}_{\rm ab}^{~2}-\frac{Gm_{\rm a}m_{\rm b}}{r_1} \\
E_{\rm B2} =& \frac{1}{2}\mathcal{M}_2\dot{\vec{r}}_{\rm cd}^{~2}-\frac{Gm_{\rm c}m_{\rm d}}{r_2} \notag.
\end{align}
In the above equations, $\vec{r}_{\rm ab}$ is the separation vector between mass a and mass b, $\vec{r}_{\rm cd}$ is the separation vector between mass c and mass d, and $\vec{r}_{\rm s}$ is the separation vector between the two binary centers of mass, as is shown in Fig. \ref{fig:2p2}.

There is likewise a well-defined final state angular momentum, $\vec{L}_0 = \vec{L}_{\rm s}+\vec{L}_{\rm B1} + \vec{L}_{\rm B2}$, where
\begin{align}
\vec{L}_{\rm s}=& m(\vec{r}_{\rm s} \times \dot{\vec{r}}_{\rm s}) \notag \\
\vec{L}_{\rm B1}=& \mathcal{M}_1 (\vec{r}_{\rm ab} \times \dot{\vec{r}}_{\rm ab}) \\
\vec{L}_{\rm B2}=& \mathcal{M}_2 (\vec{r}_{\rm cd} \times \dot{\vec{r}}_{\rm cd}). \notag
\end{align}
The two final state binaries in this outcome have characteristic semi-major axes $a_1$ and $a_2$.  We choose $a_1 > a_2$.  If $a_1 \gg a_2$, then the 2+2 problem reduces to the standard Monaghan 2+1 formalism with extra degrees of freedom.  More specifically, we can apply the loss cone formalism while treating the second binary as a point particle (although we must account for its reservoir of energy and angular momentum).

Operating first in this limit, we simplify further by working also in the low angular momentum limit.  In this case, the density of escape configurations per unit energy is
\begin{equation}
\sigma = \int ...\int \delta \left(E_{\rm s} + E_{\rm B1} + E_{\rm B2} - E_0 \right) {\rm d}\vec{r}_{\rm s}{\rm d}\vec{p}_{\rm s}{\rm d}\vec{r}_{1}{\rm d}\vec{p}_{1}{\rm d}\vec{r}_{2}{\rm d}\vec{p}_{2}.
\end{equation}
We eliminate three variables of integration with the following simplification:
\begin{align}
&\int \int \int \delta \left(\frac{p_{\rm s}^2}{2m}-\frac{Gm_{\rm B1}m_{\rm B2}}{r_{\rm s}} +E_{\rm B1} + E_{\rm B2} - E_0\right){\rm d}\vec{p}_{\rm s} \notag \\
&=4\pi f_{\rm LC} \int_0^\infty \delta \left(\frac{p_{\rm s}^2}{2m}-\frac{Gm_{\rm B1}m_{\rm B2}}{r_{\rm s}} +E_{\rm B1} + E_{\rm B2} - E_0\right) p_{\rm s}^2 {\rm d}p_{\rm s} \notag \\
&= 4\pi f_{\rm LC} m \sqrt{2m\left(E_0 + \frac{Gm_{\rm B1}m_{\rm B2}}{r_{\rm s}}-E_{\rm B1}-E_{\rm B2} \right)}.
\end{align}
Under the assumption that $a_1 \gg a_2$, we use the areal loss cone factor
\begin{equation}
f_{\rm LC} = \frac{\alpha^2 a_1^2}{4r_{\rm s}^2}.
\end{equation}
Here $\alpha$ is a dimensionless fudge factor of order unity that ultimately must be calibrated from numerical scattering experiments.  In the classic 2+1 problem, $\alpha \approx 7$.

We next evaluate
\begin{align}
&\int \int\int \frac{2^{1/2}\pi \alpha^2 a_1^2 m^{3/2}}{r_{\rm s}^{2}}\sqrt{E_0-\frac{Gm_{\rm B1}m_{\rm B2}}{r_{\rm s}}-E_{\rm B1}-E_{\rm B2}}{\rm d}\vec{r}_{\rm s} \notag \\
&= 2^{5/2}\pi^2 \alpha^2 m^{3/2}a_1^2 \int_0^R\sqrt{E_0+\frac{Gm_{\rm B1}m_{\rm B2}}{r_{\rm s}}-E_{\rm B1}-E_{\rm B2}}{\rm d}r_{\rm s} \notag \\
& \approx 2^{7/2}\pi^2\alpha^2m^{3/2}a_1^2\sqrt{Gm_{\rm B1}m_{\rm B2}}.
\end{align}
In the last approximate equality, we have taken $R \lesssim 3 a_1$.  
The density of states for this outcome is now
\begin{align}
\sigma \approx &2^{3/2}\pi^2\alpha^2m^2(GMR)^{1/2}(Gm_{\rm a}m_{\rm b})^2 \notag \\
&\times \int...\int \frac{{\rm d}\vec{r}_{\rm ab}{\rm d}\vec{p}_{\rm ab}{\rm d}\vec{r}_{\rm cd}{\rm d}\vec{p}_{\rm cd}}{|E_{\rm B1}|^2}.
\end{align}
Further simplification (making use of isotropy) gives
\begin{align}
\sigma \approx ~ & 4a^{2}\pi ^{5} (Gm_{a}m_{b})^{7/2} R^{1/2}m_{B}^{3/2}M^{-3/2}m_{e}^{2} \notag \\ 
& \times \int\int \frac{{\rm d}E_{\rm B1}{\rm d}E_{\rm B2}}{|E_{\rm B1}|^{7/2}|E_{\rm B2}|^{3/2}}L_1L_2{\rm d}L_1{\rm d}L_2 \label{eq:sigma2p2a} \\
\sigma \approx ~ & 2a^{2}\pi ^{5} (Gm_{a}m_{b})^{11/2} R^{1/2}m_{B}^{3/2}M^{-3/2}m_{e}^{2} \mathcal{M} \notag \\ 
& \times \int\int \frac{{\rm d}E_{\rm B1}{\rm d}E_{\rm B2}}{|E_{\rm B1}|^{9/2}|E_{\rm B2}|^{5/2}}e_1e_2{\rm d}e_1{\rm d}e_2. \label{eq:sigma2p2b}
\end{align}

It is important to mention that in this derivation we have called m$_{e}$ the mass of the object that we consider is the one that escapes (analogous to the three-body case in which the escaper is a single), which in this case is a binary (so m$_{B1}$ $=$ m$_{B2}$ $=$ m$_{e}$). 
For simplicity we will define A $:=$ $2a^{2}\pi ^{5} (Gm_{a}m_{b})^{11/2} R^{1/2}m_{B}^{3/2}M^{-3/2}m_{e}^{2} \mathcal{M}$, however it is important to remember that the value of m$_{e}$ and $\mathcal{M}$ will be different for each of the three different outcomes.\\
Eq. \ref{eq:sigma2p2b} encodes the final probability distribution of binary energies (Eq. \ref{eq:sigma2p2a} does not because its limits of integration depend on $E_{\rm B1}$ and $E_{\rm B2}$).  Specifically, these probability distributions are

\begin{align}
P(|E_{\rm B1}|){\rm d}|E_{\rm B1}|= A \frac{7}{2}|E_0|^{7/2}|E_{\rm B1}|^{-9/2}{\rm d}|E_{\rm B1}| \\
P(|E_{\rm B2}|){\rm d}|E_{\rm B2}|= A \frac{3}{2}|E_0|^{3/2}|E_{\rm B2}|^{-5/2}{\rm d}|E_{\rm B2}|.
\end{align}
However, in this simple derivation, we have followed \citep{Valtonen&Karttunen06}, and have neglected angular momentum conservation, rendering our results inaccurate.  We now consider the power-law ansatz of \citep{Valtonen&Karttunen06},
\begin{align}
P(|E_{\rm B1}|){\rm d}|E_{\rm B1}|=~ & 2a^{2}\pi ^{5} (Gm_{a}m_{b})^{11/2} R^{1/2}m_{B}^{3/2}M^{-3/2}m_{e}^{2} \mathcal{M} \notag \\ 
& \times
(n-1)|E_0|^{n-1}|E_{\rm B1}|^{-n}{\rm d}|E_{\rm B1}| \\
P(|E_{\rm B2}|){\rm d}|E_{\rm B2}|= ~ & 2a^{2}\pi ^{5} (Gm_{a}m_{b})^{11/2} R^{1/2}m_{B}^{3/2}M^{-3/2}m_{e}^{2} \mathcal{M} \notag \\ 
& \times
(\nu -1)|E_0|^{\nu -1}|E_{\rm B2}|^{-\nu}{\rm d}|E_{\rm B2}|.
\end{align}
In the zero angular momentum limit of the chaotic three-body problem, $n=3$, leading us to speculate that here $n=3$ and $\nu=1$.  This, however, would imply a UV divergence in $P(|E_{\rm B2}|)$ as $E_{\rm B2} \rightarrow \infty$.  We fix this by truncating at a maximum energy $E_{\rm max}$ motivated by physics beyond the Newtonian point particle limit of this paper (e.g. physical collisions, tidal interactions, relativistically unstable orbits, etc.).  Thus, 

\begin{align}
P(|E_{\rm B2}|){\rm d}|E_{\rm B2}|= ~ & 2a^{2}\pi ^{5} (Gm_{a}m_{b})^{11/2} R^{1/2}m_{B}^{3/2}M^{-3/2}m_{e}^{2} \mathcal{M} \notag \\ 
& \times
|E_{\rm B2}|^{-1}\ln^{-1}\left(\frac{E_{\rm max}}{E_0/2} \right){\rm d}|E_{\rm B2}|.
\end{align}
These approximations will begin to break down if $a_1 \approx a_2$.

\section{The 3+1 Outcome}
\label{sec:3p1}

In Paper I, we demonstrated that to a good approximation, the distribution of escaper velocities in the 3+1 outcome can be computed with a straightforward application of the 2+1 Monaghan formalism, considering only the binding energy of the inner binary $E_{\rm B}$.  This approach works because the total binding energy, $E_{1}$, of a hierarchically stable triple (at least in the equal mass case considered in Paper I) is dominated by that of the inner binary.  Unfortunately, this approach says little about the properties of the outer orbit of the resulting triple, and this adaptation of the Monaghan formalism is only applicable to the inner binary and the escaping single star.

\begin{figure}
\includegraphics[width=85mm]{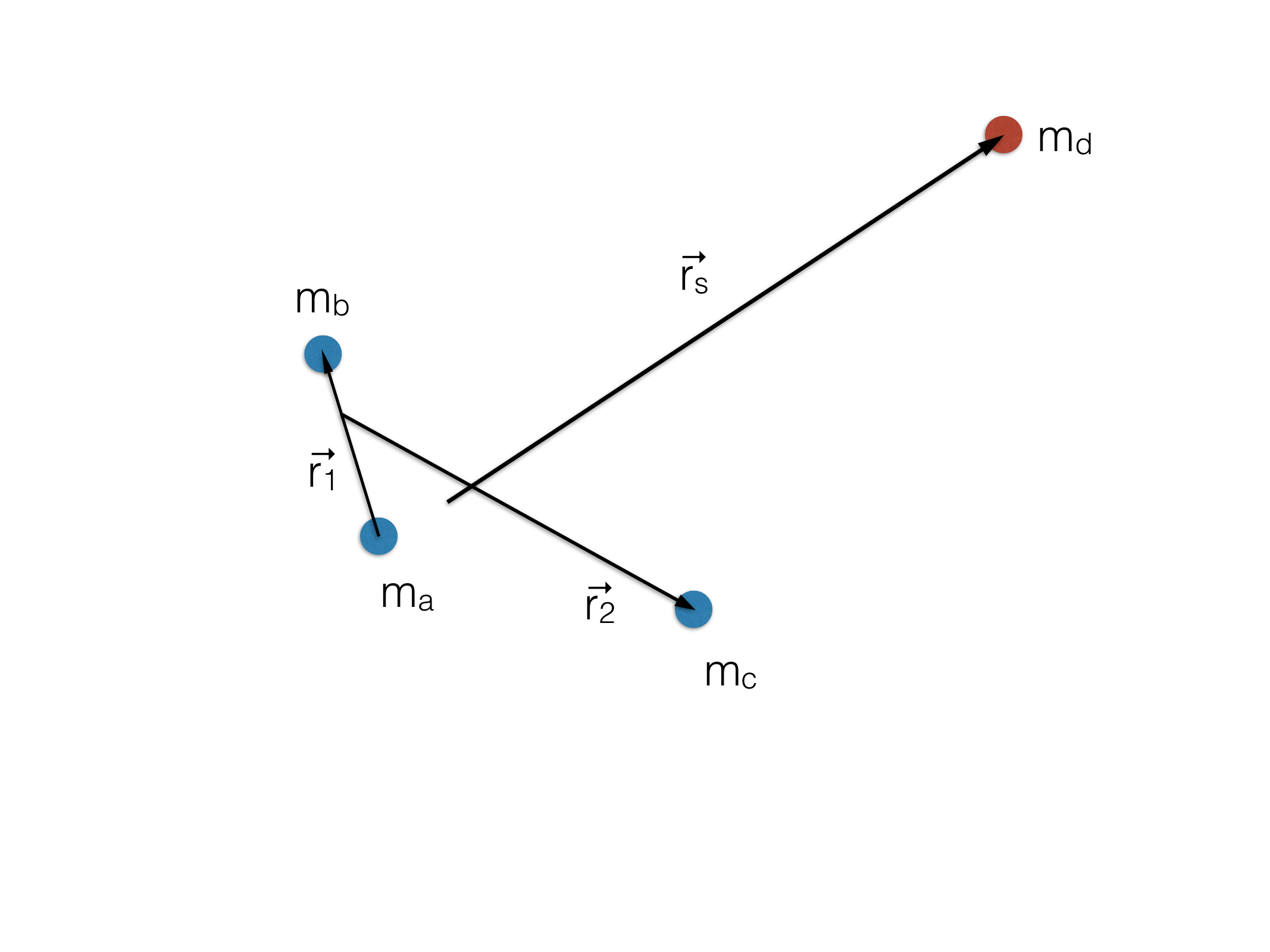}
\caption{The configuration of the 3+1 outcome.}
\label{fig:3p1}
\end{figure}

In other words, $|E_{1}|-|E_{\rm B}| \equiv |E_{\rm T}| \ll |E_{1}|$.  A cartoon sketch of the 3+1 outcome is shown in Fig. \ref{fig:3p1}.  Masses $m_{\rm a}$ and $m_{\rm b}$ form the inner binary component of the stable triple, while $m_{\rm c}$ is the outer tertiary component.  The mass $m_{\rm d}$ is escaping from the system on an unbound trajectory.  We define additional masses $m_{\rm B}=m_{\rm a}+m_{\rm b}$, $m_{\rm T}=m_{\rm B}+m_{\rm c}$, $\mathcal{M}_{\rm B}=m_{\rm a}m_{\rm b}/m_{\rm B}$, and $\mathcal{M}_{\rm T}=m_{\rm B}m_{\rm c}/m_{\rm T}$.

\subsection{Strongly Hierarchical Triples}

When masses are comparable, the stable triple produced in the 3+1 outcome can be thought of in the following way: the inner binary contains the bulk of the energy $E_1$, while the outer binary contains the bulk of the angular momentum $\vec{L}_1$.  In paper I, we showed that the standard 2+1 Monaghan formalism applies reasonably well to the binding energy distributions,  
so now applying factor A (with their respective values) we found 
\begin{equation}
P(|E_{\rm B}|){\rm d}|E_{\rm B}|=A(n-1)|E_0|^{n-1}|E_{\rm B}|^{-n}{\rm d}|E_{\rm B}|,
\end{equation}
where $n=9/2$ if angular momentum conservation is neglected and $n=3$ in zero-angular momentum ensembles.  We can determine the distribution of outer triple binding energies at a similar level of approximation by assuming a thermal distribution of outer eccentricities $e_{\rm T}$, i.e. ${\rm d}N/{\rm d}e_{\rm T}=2e_{\rm t}$.  Then, under the assumption that the outer orbit angular momentum $L_{\rm T} \gg L_{\rm B}$, we use the relation $e_{\rm T}^2=1-L_{\rm T}^2/(\mathcal{M}_{\rm T}^2Gm_{\rm T}a_{\rm T})$ to compute
\begin{equation}
\frac{{\rm d}N}{{\rm d}a_{\rm T}} = \frac{{\rm d}N}{{\rm d}e_{\rm T}}\frac{{\rm d}e_{\rm T}}{{\rm d}a_{\rm T}} = \frac{L_{\rm T}^2}{\mathcal{M}_{\rm T}^2Gm_{\rm T}a_{\rm T}^2},
\end{equation}
or, equivalently,
\begin{equation}
\frac{{\rm d}N}{{\rm d}E_{\rm T}} = \frac{2L_{\rm T}^2}{\mathcal{M}_{\rm T}^2G^2m_{\rm T}m_{\rm B}m_{\rm c}}. \label{eq:3p1outerE}
\end{equation}
For fixed $L_{\rm T}$, Eq. \ref{eq:3p1outerE} specifies the distribution of $|E_{\rm T}|$, which varies from $0$ to a maximum value $\lesssim |E_1|/2$.

We can proceed further under the assumption of small angular momentum in the inner binary; in this limit, the angular distribution of $\vec{L}_{\rm B}$ will be approximately isotropic with respect to $\vec{L}_{\rm T}$.  If we define a reference axis $\hat{z} \parallel \vec{L}_{\rm B}$, then the distribution of misalignment angles 
\begin{equation}
\frac{{\rm d}N}{{\rm d}\sin \theta} = \frac{1}{2},
\end{equation}
where $\cos\theta \equiv \hat{L}_1 \cdot \hat{L}_{\rm B} = L_1^{\rm z}/L_1$.  In the remainder of this section, we denote the $z$ component of a vector with a superscript $z$, and the components orthogonal to $\hat{z}$ with a superscript $\perp$.

We now complete this perturbative calculation: having assumed a distribution of $\theta$ which is isotropic, we wish to know the distribution of a different misalignment angle, $\cos \psi \equiv \hat{L}_{\rm B} \cdot \hat{L}_{\rm T} = L_{\rm T}^{\rm z}/L_{\rm T}$.  In general, $\psi \approx \theta$, but we aim here to quantify the leading order deviation from isotropy in $\psi$.  Since $\vec{L}_1 = \vec{L}_{\rm B} + \vec{L}_{\rm T}$, we can write $L_1^{\rm z} = L_{\rm B} + L_{\rm T}^{\rm z}$ and $L_1^\perp = L_{\rm T}^\perp$.  This yields
\begin{equation}
\cos \psi = \frac{L_1\cos\theta - L_{\rm B}}{(L_1^2+L_{\rm B}^2-2L_1L_{\rm B}\cos\theta)^{1/2}},
\end{equation}
and the quadratic formula then provides 
\begin{equation}
\cos\theta = \frac{L_{\rm B}}{L_{\rm 0}}\sin^2\psi \pm \cos\psi \sqrt{1-\frac{L_{\rm B}^2}{L_1^2}\sin^2\psi}. \label{eq:thetaOfPsi}
\end{equation}
The final distribution of interest is ${\rm d}N/{\rm d}\psi = ({\rm d}N/{\rm d}\theta)({\rm d}\theta/{\rm d}\psi) $, which evaluates to 
\begin{equation}
\frac{{\rm d}N}{{\rm d}\psi} = \frac{1}{2}\cos\theta \left( \frac{{\rm d}\psi}{{\rm d}\theta} \right)^{-1},
\end{equation}
where 
\begin{equation}
\frac{{\rm d}\psi}{{\rm d}\theta} = L_1^2\sin\theta \csc\psi \frac{L_1 - L_{\rm B}\cos\theta}{(L_1^2+L_{\rm B}^2-2L_1L_{\rm B}\cos\theta)^{3/2}},
\end{equation}
and both $\cos \theta$ and $\sin \theta$ can be computed from Eq. \ref{eq:thetaOfPsi}.

\section{The 2+1+1 Outcome}
\label{sec:2p1p1}

\begin{figure}
\includegraphics[width=85mm]{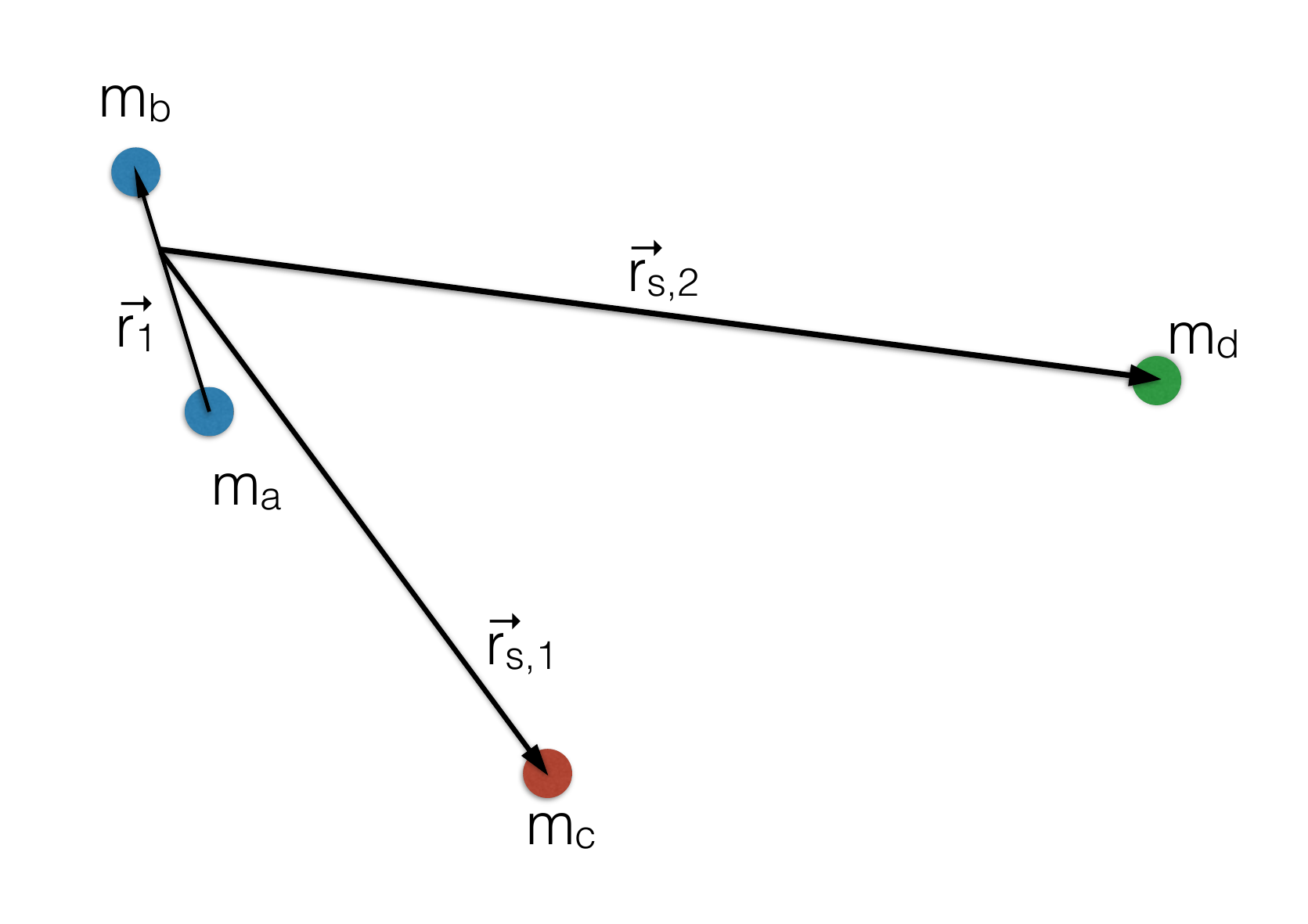}
\caption{The configuration of the 2+1+1 outcome.}
\label{fig:2p1p1}
\end{figure}

The 2+1+1 case is in some ways the most distinct from the classic 2+1 problem.  It differs not only in additional degrees of freedom, but more fundamentally in its complicated causality.  The 2+2, 3+1, and 2+1 scenarios terminate a single impulsive escape, but the 2+1+1 does not, and two different escape events must be considered.  In Fig. \ref{fig:2p1p1}, we show a cartoon of the 2+1+1 outcome.  Here masses $m_{\rm a}$ and $m_{\rm b}$ form the survivor binary, while masses $m_{\rm c}$ and $m_{\rm d}$ are escaping on unbound orbits.  The order of escape matters: the distribution of parameters for the first escaper ($m_{\rm c}$) will differ from the parameter distribution for the second escaper ($m_{\rm d}$).

We begin by making an approximation of {\it sequential escape}: we assume that in general, a metastable triple is formed after the escape of particle C, and that particle D is only ejected after the gravitational influence of C becomes negligible.  This approximation lets us apply the standard 2+1 Monaghan formalism in an iterated way.  Based on the numerical scattering experiments of Paper I, we believe it to be well justified for low virial ratios ($k\ll 1$) but a poor approximation for high virial ratios ($k\approx 1$), when both ejected stars are ejected almost simultaneously.  We first estimate the distribution of binding energies $E_{\rm T}$ of the metastable triple:
\begin{equation}
P(|E_{\rm T}|){\rm d}|E_{\rm T}|= \frac{7}{2}|E_{\rm 0}|^{7/2}|E_{\rm T}|^{-9/2}{\rm d}|E_{\rm T}|. \label{eq:metastableTriple}
\end{equation}
As before, $E_0$ is the conserved total energy of the four-body encounter.  For a given $E_{\rm T}$ value, we can take the standard 2+1 distribution of binding energies for $E_{\rm B}$ (the binding energy of the final surviving binary), but if we want a distribution of $E_{\rm T}$ values, we need to integrate over Eq. \ref{eq:metastableTriple}:
\begin{align}
P(|E_{\rm B}|){\rm d}|E_{\rm B}| =& \int_{|E_0|}^{|E_{\rm B}|}\frac{7}{2}|E_{\rm T}|^{7/2}|E_{\rm B}|^{-9/2}P(|E_{\rm T}|){\rm d}|E_{\rm T}| {\rm d}|E_{\rm B}|   \notag \\
=& \frac{49}{4}|E_0|^{7/2}|E_{\rm B}|^{-9/2}\ln (E_{\rm B}/E_0).
\end{align}
More generally, if we substitute a power law index $n$ for the triple binding energy distribution ($9/2$ above) we find
\begin{equation}
P(|E_{\rm B}|){\rm d}|E_{\rm B}| = A (n-1)^2|E_0|^{n-1}|E_{\rm B}|^{-n}\ln (E_{\rm B}/E_0).
\end{equation}
Likewise, we can apply the results of the standard 2+1 formalism.

\section{Branching Ratios}
\label{sec:branching}
 
The "branching ratio" defines the probability of obtaining a given outcome, for a given total encounter energy and angular momentum.  Hence, for the chaotic four-body problem in the point-particle limit with total energy E $<$ 0, there are three branching ratios to consider.  In general, the relative fractions for these different outcomes must be determined using numerical scattering simulations.  However, as we are about to show, these branching ratios can also be computed analytically, if all particles are identical.  

Consider performing $N_{\rm 0}$ simulations of a chaotic four-body interaction involving identical point-particles, with nearly identical initial conditions.  Given $N_{\rm 0}$ simulations, we must obtain $N_{\rm 0}$ outcomes.  Then, the total number of simulations performed can be written:
\begin{equation}
\label{eqn:N0}
N_{\rm 0} = N_{\rm 2+1+1} + N_{\rm 3+1} + N_{\rm 2+2},
\end{equation}   
where $N_{\rm 2+1+1}$, $N_{\rm 3+1}$ and $N_{\rm 2+2}$ correspond to the number of simulations resulting in, respectively, the 2+1+1, 3+1 and 2+2 outcomes.  Now, by conservation of energy, we must also find that the total amount of energy put in across all simulations is equal to the total energy we get back out.  In other words:
\begin{align}
\label{eqn:energyconserv}
\begin{split}
    N_{\rm 0}E_{\rm 0} = & N_{\rm 3+1}\int E\frac{dN_{\rm 3+1}}{dE}dE + N_{\rm 2+1+1}\int E\frac{dN_{2+1+1}}{dE}dE\\
    & + N_{\rm 2+2} \int E \frac{dN_{2+2}}{dE}dE
\end{split}
\end{align}
Similarly, the total angular momentum must be conserved, ensuring that the following must be true:
\begin{equation}
\label{eqn:momconserv}
\begin{split}
    N_{\rm 0}L_{\rm 0} = & N_{\rm 3+1}\int L\frac{dN_{\rm 3+1}}{dL}dL + N_{\rm 2+1+1}\int L\frac{dN_{2+1+1}}{dL}dL\\
    & + N_{\rm 2+2}\int L\frac{dN_{2+2}}{dL}dL
\end{split}
\end{equation}
Each term on the right-hand-sides of Equations~\ref{eqn:energyconserv} and~\ref{eqn:momconserv} must be broken up in to the individual contributions from each decay product (i.e., single, binary and/or triple star(s)).  The limits of the resulting integrals must then be chosen appropriately.  For example, for the 2+1+1 outcome, we have:
\begin{equation}
\label{eqn:energyconserv2}
\begin{split}
    \int E\frac{dN_{2+1+1}}{dE}dE = & \int E_{\rm S,1}\frac{dN_{2+1+1}}{dE_{\rm S,1}}dE_{\rm S,1}\\
    & +  \int E_{\rm S,2}\frac{dN_{2+1+1}}{dE_{\rm S,2}}dE_{\rm S,2}\\
    & +  \int E_{\rm B}\frac{dN_{2+1+1}}{dE_{\rm B}}dE_{\rm B}  
\end{split}
\end{equation}

\begin{align}
\label{eqn:energyconserv3}
\begin{split}
    \int E\frac{dN_{2+1+1}}{dE}dE = & \frac{1}{2}m_{\rm S,1}\int_0^{\infty} f(E_{\rm S,1})v_{\rm S,1}^2dE_{\rm S,1}\\
    & + \frac{1}{2}m_{\rm S,2}\int_0^{\infty} f(E_{\rm S,2})v_{\rm S,2}^2dE_{\rm S,2}\\ 
    & + 
    \int_{-\infty}^0 f(E_{\rm B})E_{\rm B}dE_{\rm B},
\end{split}
\end{align}
where the indices $1$ and $2$ correspond to, respectively, the first and second ejected single stars, and all distributions correspond to those presented for the 2+1+1 outcome.  Note as well that $f(E_{\rm S,i}) = f(v_{\rm S,i})/(m_{\rm S,i}v_{\rm S,i})$.

If we divide both sides of all three equations by $N_{\rm 0}$, then Equations~\ref{eqn:N0}, ~\ref{eqn:energyconserv} and~\ref{eqn:momconserv} constitute three equations, each with the same three unknowns.  Hence, this system of equations is solvable.  The factor in front of each term corresponds to the branching ratio for that outcome.     

We caution that if the particles are not identical, then the formalism presented here for calculating branching ratios for the different outcomes is no longer valid, strictly speaking.  We defer this issue to a future paper, along with a more thorough comparison between the predicted branching ratios and the results of numerical scattering simulations.

\section{Methods} \label{sec:methods}

In this section, we describe and present the numerical scattering simulations used to test directly the analytic distribution functions derived in the preceding sections.

\subsection{Numerical scattering simulations} \label{sec:sims}

The numerical scattering simulations used throughout this paper are the same as presented in \citet{Leigh+16}.  For completeness, we repeat our description of the code and initial set-up here.

We calculate the outcomes of a series of binary-binary (2+2) encounters using the \texttt{FEWBODY} numerical 
scattering code\footnote{For the source code, see http://fewbody.sourceforge.net.}.  The code integrates the usual 
$N$-body equations in configuration- (i.e. position-) space in order to advance the system forward in time, using the 
eighth-order Runge-Kutta Prince-Dormand integration method with ninth-order error estimate and adaptive time-step.  
For more details about the \texttt{FEWBODY} code, we refer the reader to \citet{fregeau04}.  

{The outcomes of these 2+2 encounters are studied for the initial virial ratio $k$ $=$ 0, where $k$ is defined as:}
\begin{equation}
\label{eqn:virial}
k = \frac{T_{\rm 1} + T_{\rm 2}}{E_{\rm b,1} + E_{\rm b,2}},
\end{equation}
here the indexes 1 and 2 correspond to the two initial binaries.  The initial kinetic energy corresponding to the 
centre of mass motion of binary $i$ is:
\begin{equation}
\label{eqn:kinetic}
T_{\rm i} = \frac{1}{2}m_{\rm i}v_{\rm inf,i}^2,
 \end{equation}
where $m_{\rm i} =$ $m_{\rm i,a} + m_{\rm i,b}$ is the total binary mass and $v_{\rm inf,i}$ is the initial 
centre of mass velocity for binary $i$.  The initial orbital energy of binary $i$ is:
\begin{equation}
\label{eqn:orbital} 
E_{\rm b,i} = -\frac{Gm_{\rm i,a}m_{\rm i,b}}{2a_{\rm i}},
\end{equation}
where $m_{\rm i,a}$ and $m_{\rm i,b}$ are the masses of the binary components and $a_{\rm i}$ is the initial 
orbital separation.  Given this definition for the virial ratio, $k =$ 0 corresponds to the binaries starting from rest, and maximizes the fraction of longer-lived chaotic interactions (which is a necessary prerequisite to apply the Monaghan formalism).

All objects are point particles with masses of 1 M$_{\odot}$.  All binaries have $a_{\rm i} =$ 1 AU initially, and 
eccentricities $e_{\rm i} =$ 0.  We fix the impact parameter at $b =$ 0 for all simulations.  
The angles defining the initial relative configurations of the binary orbital planes and phases are 
chosen at random.  

We use the same criteria as \citet{fregeau04} to decide when a given encounter is complete.  To first order, this is defined as 
the point at which the separately bound hierarchies that make up the system are no longer interacting with each other or 
evolving internally.  More specifically, the integration is terminated when the top-level hierarchies have positive relative 
velocity and the corresponding top-level $N$-body system has positive total energy.  Each hierarchy must also be dynamically 
stable and experience a tidal perturbation from other nodes within the same hierarchy that is less than the critical value 
adopted by \texttt{FEWBODY}, called the tidal tolerance parameter.  For this study, we adopt the tidal tolerance parameter 
$\delta =$ 10$^{-7}$ for all simulations.\footnote{The more stringent the tidal tolerance parameter is chosen to be, the closer to a 
"pure" $N$-body code the simulation becomes.}  This choice for $\delta$, while computationally expensive, is needed to maximize 
the accuracy of our simulations, and ensure that we have converged on the correct encounter outcome (see \citealt{geller15} 
for more details).

Because of the isotropy of our initial conditions, the typical four-body encounter we simulate has $L_0 > 0$.  If one considers a binary-binary scattering event where the two initial binaries have isotropically oriented angular momentum vectors of magnitude $L_1$ and $L_2$ ($L_1 \ge L_2$ by assumption), then the total angular momentum $\vec{L}_0 = \vec{L}_1 + \vec{L}_2$ spans a range of magnitudes from $L_1-L_2$ to $L_1+L_2$ with a distribution
\begin{equation}
\frac{{\rm d}N}{{\rm d}L_0} = \frac{L_0}{2L_1 L_2}.
\end{equation}
The first moment of this distribution is
\begin{equation}
\langle L_0 \rangle = L_1 + \frac{L_2}{3L_1}L_2.
\end{equation}
If we specialize now to our initial conditions (equal masses $m$, equal initial semi-major axes, $L\equiv L_1=L_2$), we find that 
\begin{equation}
\frac{\langle L_0 \rangle}{L_{\rm max}} = \frac{4\sqrt{2}}{15},
\end{equation}
where we have followed \citet{Valtonen&Karttunen06} in defining the maximum system angular momentum $L_{\rm max} \equiv \frac{5}{2}G\sqrt{m^5/|E_0|}$.  Their numerical fitting formula for the classic 2+1 problem predicts $n=3+18\tilde{L}^2$ for ensembles of resonant three-body encounters with angular momentum $\tilde{L}=L/L_{\rm max}$.  This gives us a naive expectation of $n \approx 5.6$ for our numerical simulations.

\section{Results} \label{sec:results}
In this section, we compare the results of our numerical scattering simulations to the fitting formulae presented in the previous sections.

\subsection{Comparing to the simulations} \label{sec:compare} 
\begin{figure}
\includegraphics[width=85mm]{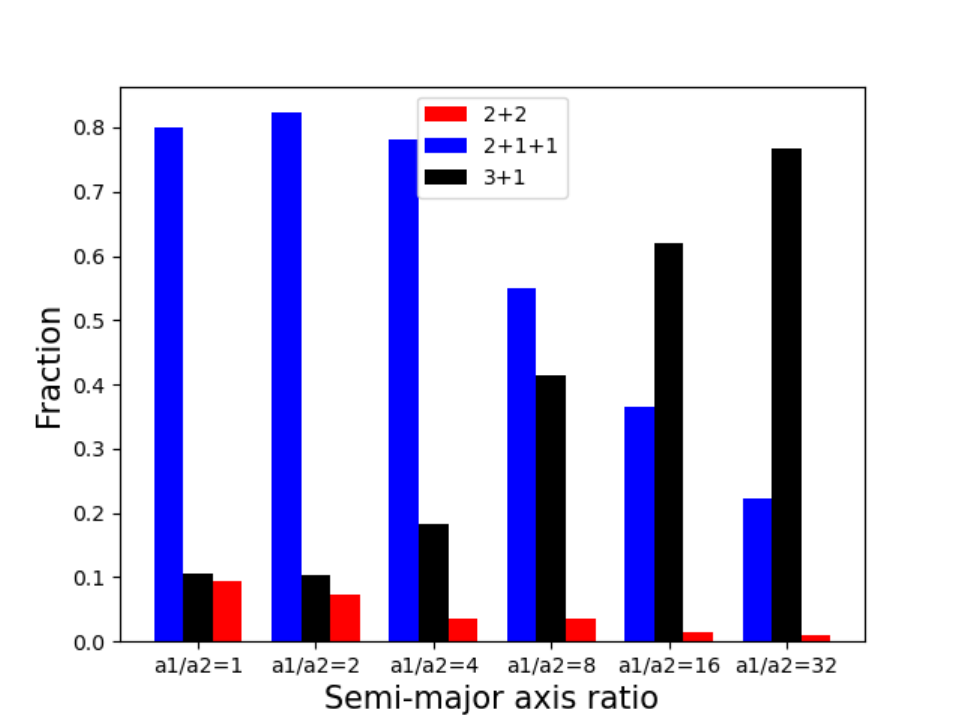}
\caption{{{The fraction of different outcomes for our binary-binary scattering simulations as a function of the ratio of the initial semi-major axes of the binaries, or a$_{\rm 1}$/a$_{\rm 2}$. We vary the ratio as a$_{\rm 1}$/a$_{\rm 2}$ $=$ [1, 2, 4, 8, 16, 32].} 
}
\label{fig:fig0}}
\end{figure}

{In Figure~\ref{fig:fig0}, we present the final outcome distributions after the interaction between our two initial binaries. We separate our results for different semi-major axis ratios
(a$_{1}$/a$_{2}$, indicated on the x-axis) and show the fraction of simulations resulting in each of our three possible outcomes (i.e., 3+1, 2+1+1 and 2+2).  For each combination of a$_{1}$/a$_{2}$ we perform 10,000 scattering simulations. The results for the 2+2, 3+1 and 2+1+1 cases are shown in red, black and blue colour bars, respectively. Note that this colour relationship will be used for all figures in this paper.}  We note that these branching ratios should be computable explicitly using the equations in Section~\ref{sec:branching}, but we defer this to a future paper, since we first need to re-derive the analytic functions in this paper from first principles as in \citet{stoneleigh19} or \cite{GinatPerets2021} to obtain the needed angular momentum dependences.

We see that for the case in which our initial semi-major axes are relatively similar (1 $\leq$ a$_{1}$/a$_{2}$ $\leq$ 4) the largest outcome fraction corresponds to the 2+1+1 scenario. This is in agreement with what was shown by \citet{mikkola83} and \citet{Leigh+16} for identical initial conditions. The trend changes as the ratio of the semi-major axis increases. For a$_{1}$/a$_{2}$ $>$ 8 the scenario that occurs the most corresponds to the 3+1 case, that is a particle of the system is transformed into an escaper leaving behind a dynamically stable triple system. Finally, the formation of the 2+2 case is always the least probable, tending to decrease as the semi-major axis ratio increases. This is because for this to occur, both binaries must form simultaneously, which effectively requires that two stars be ejected in similar directions, with similar ejection times and escape velocities, rendering this outcome improbable.  Alternatively, this can be viewed as the wider binary struggling more and more to eject the more compact binary (in analogy to the the three-body case) as the ratio of semi-major axes increases at fixed particle mass.

\begin{figure}
\includegraphics[width=85mm]{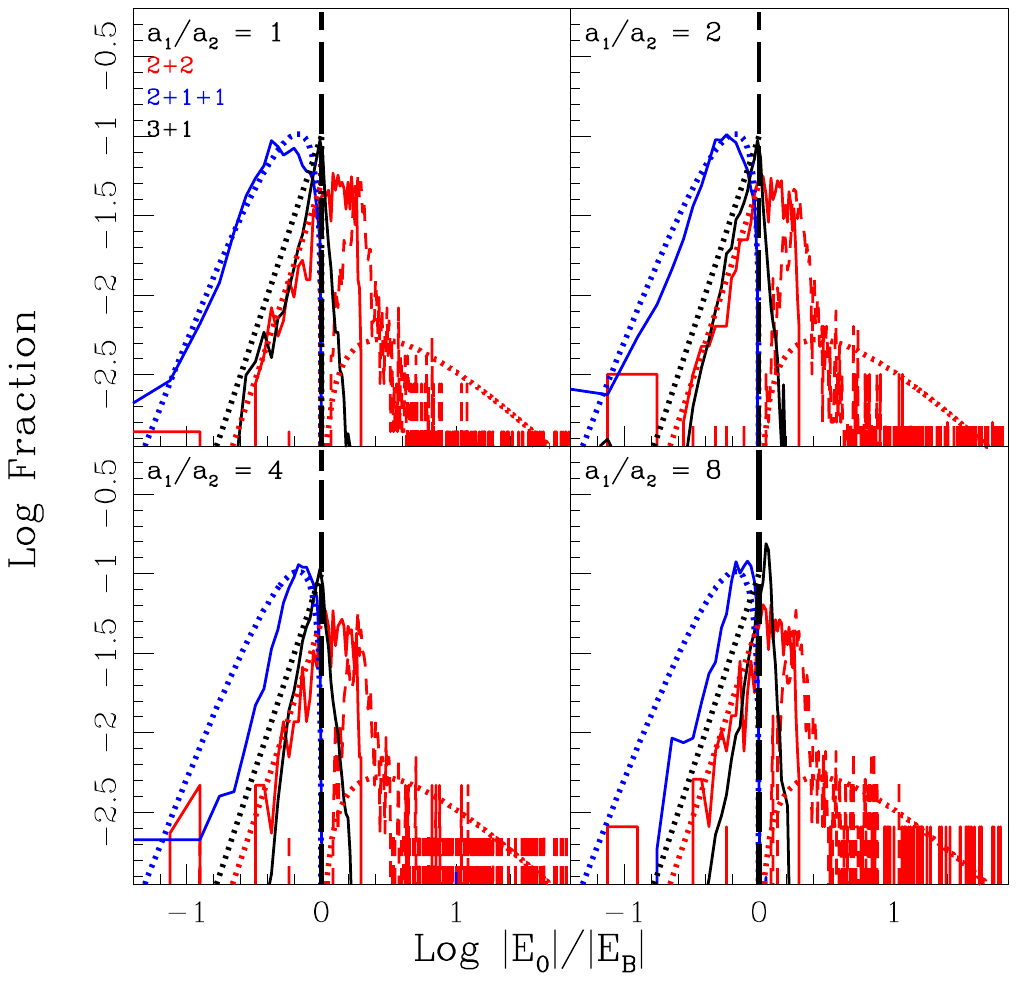}
\caption{{{The distributions of final binary binding energies are shown for each encounter outcome, parameterized using the total encounter energy E$_{\rm 0}$ as z $=$ E$_{\rm 0}$/E$_{\rm B}$. The colours are the same as in Figure~\ref{fig:fig0}.
The solid lines represent the values of the simulations while the dotted lines show the values obtained analytically.
The black vertical dashed line shows the ratio E$_{0}$/E$_{B}$ $=$ 1.
Each panel shows the distributions for a different value of the initial semi-major axis ratio where a$_{1}$/a$_{2}$ $=$ [1, 2, 4, 8].  All histograms have been normalized by the total number of simulations that resulted in the corresponding outcome.  Note that for the 2+2 case (red color) there is a solid line and another dashed line, which represent the simulated values for the compact and wide binary, respectively.}}
\label{fig:fig1}}
\end{figure}

{{In Figure~\ref{fig:fig1}, we present normalized histograms of the final distributions of 
binary orbital energies, parameterized using the total encounter 
energy E$_{0}$ as z $=$ E$_{0}$/E$_{B}$. This is the result of our binary-binary 
scattering experiments, for different values of the semi-major axis
ratio (a$_{1}$/a$_{2}$, as in Figure~\ref{fig:fig0}). For the 2+2 case, we show the orbital energies of both final binaries, divided according to their final orbital energies using the solid and dashed lines for the compact and wide binary, respectively.}  For the 3+1 case, we show only the orbital energy of the inner binary of the resulting triple.}

For the energy range between 0 $\leq$ $|$E$_{0}$$|$/$|$E$_{B}$$|$ $\leq$ 1, the analytical distributions reproduce very well what was obtained through the simulations. Especially for the 2+1+1 case, the curve is wholly reproduced, which shows that our assumption that this outcome can be modeled by applying the three-body disintegration scheme twice, by assuming additionally that each escape is well separated in time,
work quite well. The same happens for the 3+1 case, assuming that all the interaction energy is held in the inner binary of the triple system, while all the angular momentum is kept in the outer orbit of the same system. In the 2+2 case, we see that the distribution fits well for the compact binary provided $|$E$_{0}$$|$/$|$E$_{B}$$|$ $\leq$ 1, while for the case of the wider binary the distribution fits poorly. This is likely because our assumptions here begin to break down such that very few of these interactions are truly chaotic, and hence very few of our simulations should actually produce results that agree with theory.  This is supported by the fact that the good agreement between our results and theory begins to decrease as the ratio of the semi-major axis increases, particularly for a$_{1}$/a$_{2}$ $\geq$ 8. This last point for the 2+2 configuration, and the distributions for which E$_{0}$/E$_{B}$ $>$ 1, will be taken up again in Section $\ref{sec:discuss}$.

The binding energy distribution can be used to derive the escape velocity distribution $f(v_{\rm e})dv_{\rm e}$ for the escaping star(s), as given by Equation~\ref{eqn:vdist} with $m_{\rm e}$ = $m_{\rm b}$/3 = M/4 for the 3+1 case for all identical particles \citep{Leigh+16}.  This gives the following functional form (Equations 7.19 and 7.26 in \citealt{Valtonen&Karttunen06}):
\begin{equation}
\label{eqn:vdist}
f(v_{\rm e})dv_{\rm e} = \frac{((n-1)|E_{\rm 0}|^{n-1}(m_{\rm e}M/m_{\rm b}))v_{\rm e}dv_{\rm e}}{(|E_{\rm 0}| + \frac{1}{2}(m_{\rm e}M/m_{\rm b})v_{\rm e}^2)^{n}}.
\end{equation}

{{In Figure~\ref{fig:fig3} we show, for different initial a$_{\rm 1}$/a$_{\rm 2}$ ratios, the distribution of escape velocities for the single star for a 3+1 outcome (black lines), as well as for both single stars for a 2+1+1 outcome (blue lines), where the solid lines show the simulated data and the dotted lines show the analytic fits. Note that to calculate the distribution analytically using Equation~\ref{eqn:vdist}, we use the values n $=$ 4.5 to account for the angular momentum dependence and m$_{e}$ $=$ m$_{b}$/3 $=$ M/4 for the mass of the escaping particle(s).}

{{Figure~\ref{fig:fig4} shows the distributions of escape velocities for binaries for the 2+2 outcome (red lines). The solid lines show the results of our numerical scattering simulations, whereas the dotted lines show our analytic fits.  Note that by conservation of linear momentum, the escape velocity distributions are equivalent for both binaries, since we are dealing with the equal particle mass case.}}

We clearly see that the analytic distribution of the escape velocities shows a clear agreement with what was seen in the simulations. On the other hand, there is a clear tendency for poorer agreement between the theory and the simulations as the semi-major axis ratio increases (as in Figure \ref{fig:fig1}). The velocity range as well as the corresponding outcome fractions for the 3+1 and 2+1+1 cases show very similar distributions as those found in \citep{Leigh+16}, but in our case for $n = 4.5$. The same is true for the 2+2 case where our distribution behaves as expected for $n = 4.5$.

\begin{figure}
\includegraphics[width=85mm]{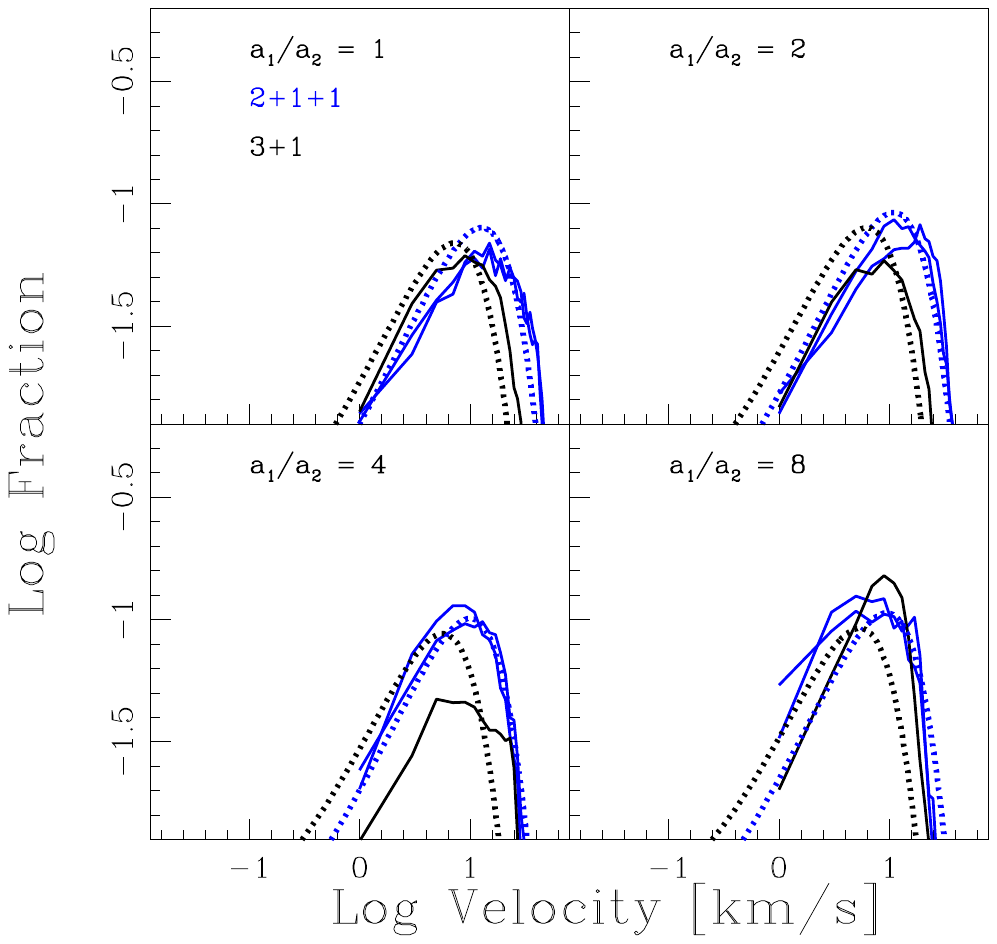}
\caption{{
Comparison between simulations and analytic results for normalized distributions of escape velocities from the single star (in km/s) for the 3+1 (black color) and 2+1+1 (blue color) outcomes.
The different insets show the same semi-major axis ratios, number of simulations, line types and colours as in Figure~\ref{fig:fig1}.
The dotted black line shows the distribution of escape velocities calculated using Equation~\ref{eqn:vdist} for a 3+1 outcome and assuming n = 4.5, corresponding to approximately isotropic scattering. The dotted blue line shows the same thing but for the 2+1+1 case assuming n = 4.5.  For both analytical curves we assume $m_{\rm e}$ = $m_{\rm b}$/3 = M/4.
Note that for the 2+1+1 case there are 2 solid blue lines, which correspond to the 2 escapers in this scenario.}
}
\label{fig:fig3}
\end{figure}}

\begin{figure}
\includegraphics[width=85mm]{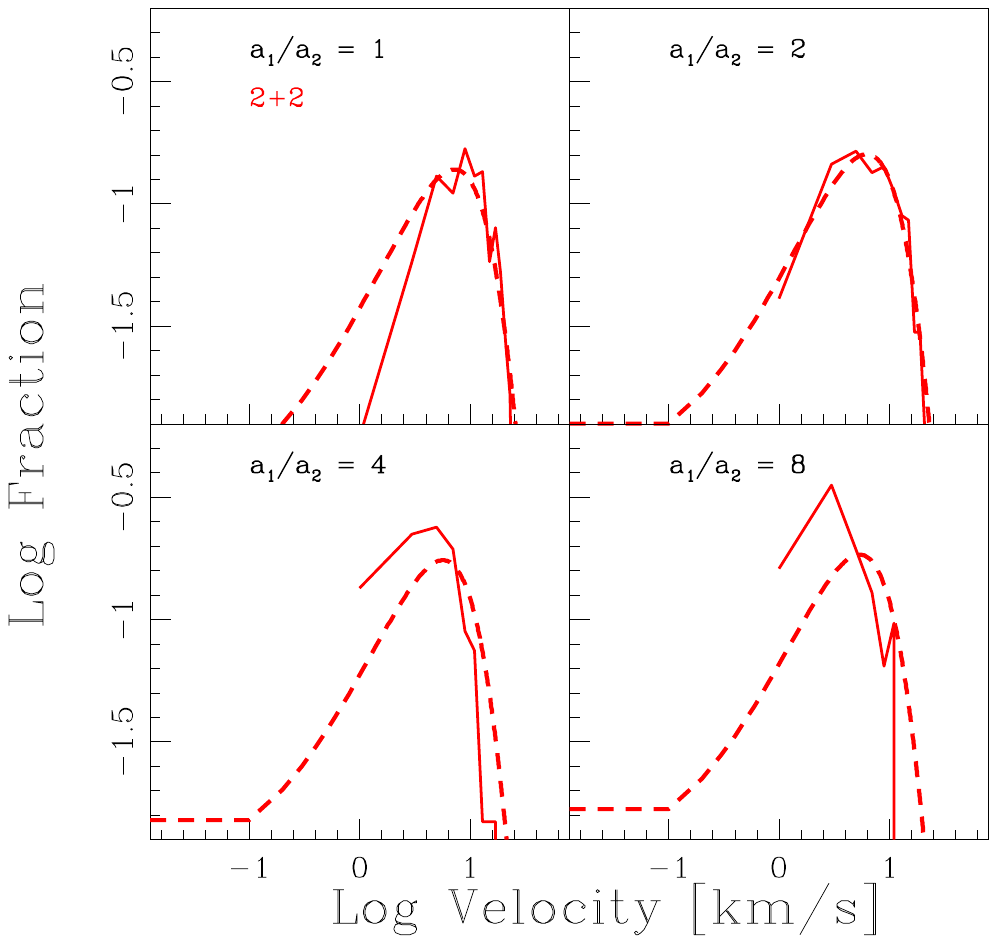}
\caption{{
The same as in Figure~\ref{fig:fig3} but for both binaries in the 2+2 scenario.
The different insets show the same semi-major axis ratios, number of simulations, line types and colours as in Figure~\ref{fig:fig1}.
The dashed red line shows the distribution of escape velocities calculated using Equation~\ref{eqn:vdist} and assuming n = 4.5, with $m_{\rm e}$ = $m_{\rm b}$ = M/2. 
Note that there is only one solid red line, since both binaries (wide and compact) have the same escape velocity distribution due to conservation of linear momentum.}
\label{fig:fig4}}
\end{figure}

\begin{figure}
\includegraphics[width=85mm]{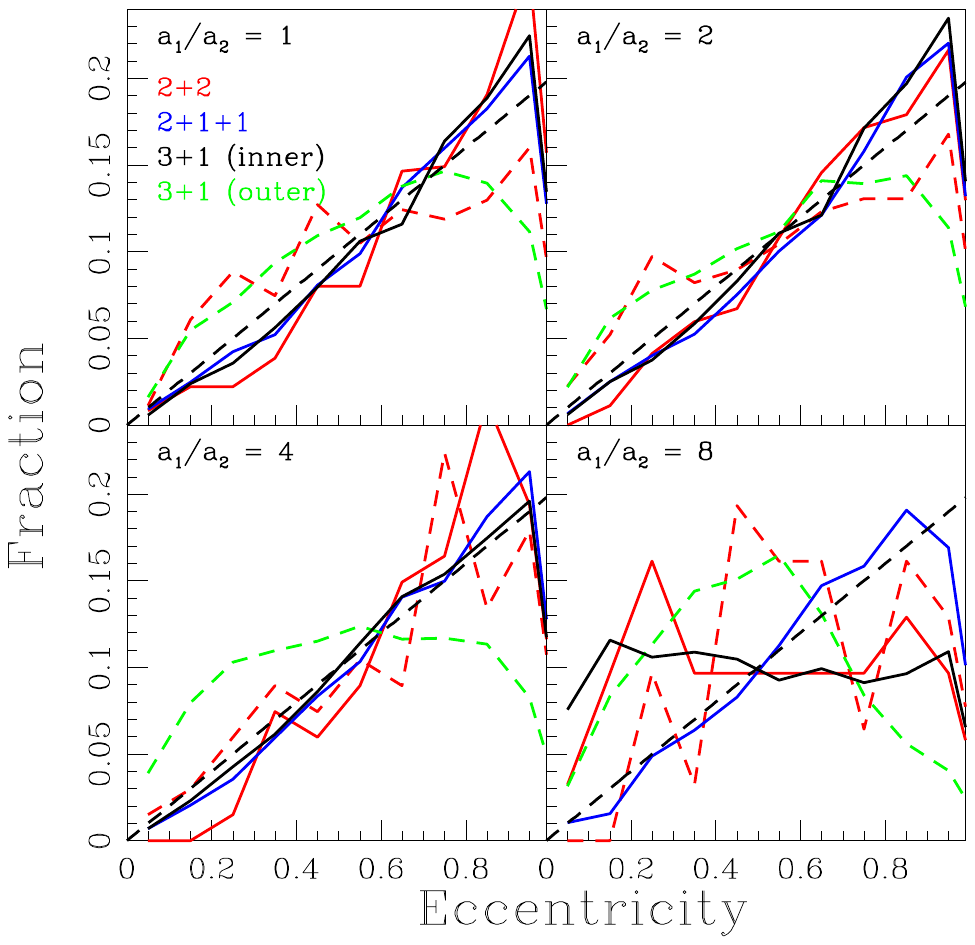}
\caption{{The distributions of final binary orbital eccentricities are shown for each encounter outcome.
The solid blue line shows the distribution of eccentricities for the binary for the 2+1+1 case.
The solid black line shows the distribution of eccentricities for the inner binary of the triple system for the 3+1 case while the dashed green line shows the same for the outer orbit of the triple system for the 3+1 case.
The red solid line shows the eccentricity distribution for the compact binary in the 2+2 case, while the red dashed line shows the distribution of eccentricities for the wide binary in the same case. 
For comparison, we plot a black dashed line showing a thermal eccentricity distribution f(e)de $=$ 2e.
The different insets show the same semi-major axis ratios and number of simulations as in Figure~\ref{fig:fig1}.
}
\label{fig:fig5}}
\end{figure}

{{In Figure~\ref{fig:fig5}, we show the final distributions of orbital eccentricities for every encounter outcome, including the inner orbits of stable triples}.  The solid black line shows a thermal eccentricity distribution f(e)de $=$ 2e, which matches the simulated data quite well for all orbits and all encounter outcomes.}   

\section{Discussion}
\label{sec:discuss}

In this section, we discuss possible caveats of our work.

First, we note that the agreement between our simulations and the analytic fits is consistently good for small initial semi-major axis ratios but becomes poorer as this ratio increases.  This is because, at least in part, a smaller fraction of the interactions become chaotic, which is a prerequisite to performing this comparison.  Hence, we are left with fewer simulations that should be in agreement with theory and must omit those simulations that ended deterministically.  Most of the interactions that cause this turn out to be simple exchanges.  In other words, in the limit of a large semi-major axis ratio, the probability that the compact binary will simply be exchanged into the wide binary becomes high, liberating one single star in the process and forming a dynamically stable hierarchical triple system.  Hence, fitting a three-body interaction, treating the more compact binary as a heavy single particle, is a more appropriate model in this limit.

In Figure \ref{fig:fig1}, we show the simulated distributions of left-over binary orbital energies for the most compact orbit in the final outcome and compare to our analytic fits.  We see good agreement between the two for E$_{\rm 0}$/E$_{\rm B} <$ 1, since beyond this limit significant angular momentum is contained in the final orbit and our analytic fits do not account for any angular momentum dependence.  For example, in the 3+1 case, the inner binary also contains angular momentum but we assume that it does not.  For this case, the angular momentum in the inner and outer orbits ultimately limits the minimum binary energy in the interior because the outer orbit cannot contain enough angular momentum to accommodate a dynamically stable hierarchical triple. This itself explains why the 2+2 distributions also tend to zero, although at lower minimum binary energies, since wider binaries can be accommodated in this case, since there is no requirement mentioned above for dynamic stability in the triplet case.

In the same figure, we can see that in the case of the wide binary in the 2+2 scenario, our approximation does not fit the simulated values.  This occurs because the observed values are given in the domain E$_{0}$/E$_{B}$ $>$ 1, which means that in this binary system there is a large amount of angular momentum. Therefore, since our analytic formalism does not explicitly take into account the angular momentum dependence, it is perhaps not surprising that we do not see good agreement between the simulated data and theory in this domain.

In Section \ref{sec:methods}, we show that we expect the analytic distributions to match the simulated ones when n $\approx$ 5.6. When performing our comparisons, we adopt a value of n $=$ 4.5 which shows the best agreement.  However, we note that our expected value of n $=$ 5.6 also does a good job of describing the data. 
 
This difference is probably due to the fact that we are not considering the total angular momentum dependence in our derivation.  Moreover, 
for our simulations we incorporated different initial semi-major axes in the four-body interaction, while for the initial derivation we assumed equal semi-major axes.
 
In Figure \ref{fig:fig5}, we see that the eccentricity distribution tends to be quite similar to the thermal distribution. While this is theoretically expected for the three-body case, assuming a detailed balance between binary creation and destruction \citep{heggie75}, we are not aware of any expectation for the four-body case. Nevertheless, the reason we see this agreement is probably the same as argued in \citep{heggie75} for the three-body case.  This is because a thermal distribution is expected for ergodized outcomes in three-body interactions, and since in this paper we treat each decomposition of the four-body case as a variation of the three-body decomposition, this distribution makes sense (e.g., the 2+1+1 case is modeled as two sequential disruptions of three-body systems).

On the other hand, we see that the distribution of eccentricities corresponding to the external component of the triple system (green dashed line) shows a distribution that does not quite match the thermal distribution for values close to 1 (highest eccentricities). Here we see a paucity of triples with high outer eccentricities relative to a thermal distribution because stable triple systems cannot exist if the external component has very large eccentricity. Otherwise, the triple system will tend to break up, ending up in the 2+1+1 configuration. As expected, this tendency is repeated in each inset (a$_{1}$/a$_{2}$ = 1, 2, 4, 8), but we see a tendency for the eccentricity distribution of the outer orbits of stable triples to flatten as the ratio of semi-major axes increases.  This is likely because we begin with all binaries being initially circular, and when there is a large ratio between their semi-major axes a simple exchange interaction is the most likely outcome, and here the outer orbit is more often left unaffected, remaining approximately circular.

\section{Conclusions}
\label{sec:conclusions}

In this paper, we have derived analytic distribution functions using the density of states formalism and an ansatz-based approach for the outcomes of four-body (i.e., binary-binary) scatterings in the equal-mass point-particle limit. We have further confronted our analytic fits with the results of numerical scattering simulations, and find good agreement.  The highlights of our results can be summarized as follows:

\begin{itemize}
\item We have derived analytic distribution functions (DFs) to describe the properties of the products of chaotic four-body interactions in the equal-mass point particle limit.   These DFs include, for the most compact orbit in the left-over binaries and/or triples in the final outcome state, the distributions of orbital energies for the left-over binaries and triples, the distributions of ejection velocities and the orbital parameters of any left-over binaries or triples.  We find good agreement between our analytic theory and the simulations for low semi-major axis ratios, since for larger ratios the angular momentum dependence would need to be integrated into our analytic formalism to expect good agreement in this limit.

\item For most of the relevant parameter space, binary-binary scatterings act to systematically destroy binaries by either forming two ejected singles or a stable hierarchical triple instead.

\item The 2+1+1 outcome (i.e., one binary and two singles are formed) tends to form the most compact binaries.

\item The 2+2 outcome (i.e., two binaries are produced) is consistently the least likely outcome for all ratios of the initial binary semi-major axes, and tends to produce the widest binaries.  This is because, in order to form two binaries in the end, effectively the more compact final binary must eject the other two stars at about the same time, in similar directions and with comparable ejection velocities.  Alternatively, this can be viewed as the wider binary having more and more difficulty in ejecting the more compact binary (in analogy to the three-body case) as the ratio of semi-major axes increases (at fixed particle mass).

\item All outcomes of binary-binary scatterings produce binaries with a distribution of eccentricities consistent with being thermal.  This is the case except for very large initial semi-major axis ratios, for which we find a flat eccentricity distribution for the inner binaries of dynamically-formed triples (while that for the 2+1+1 outcome remains consistent with thermal).  Since it is those binary-binary scatterings with the largest semi-major axis ratios that produce the most triples, we naively expect that the inner binaries of dynamically-formed triples should show an approximately flat distribution of orbital eccentricities (in part since we assume initially circular orbits).  Finally, for the outer orbits of stable triples, we see a slight deviation from a thermal distribution at high eccentricities, since here very high values for the eccentricity are forbidden if the formed triple is to be dynamically stable.

\item We have derived a prediction for the distribution of inclination angles between the inner and outer orbital planes of dynamically-formed stable hierarchical triples.  We find that it deviates from an isotropic distribution more and more with increasing angular momentum, potentially allowing for an observational signature to test if triples are primarily formed dynamically (including during the star formation phase).

\end{itemize}

\section*{Acknowledgements}
We very gratefully acknowledge discussions with Barry Ginat and Hagai Perets. CMBR acknowledges financial support from Millenium Nucleus NCN19\_058 (TITANs). NWCL gratefully acknowledges the generous support of a Fondecyt Iniciaci\'on grant 11180005 and a Fondecyt Regular grant 1230082, as well as support from Millenium Nucleus NCN19\_058 (TITANs) and funding via the BASAL Centro de Excelencia en Astrofisica y Tecnologias Afines (CATA) grant PFB-06/2007.  NWCL also thanks support from ANID BASAL project ACE210002 and ANID BASAL projects ACE210002 and FB210003. 

\section*{Data Availability}
The data underlying this article will be shared on reasonable request to the corresponding author.

\bsp	
\label{lastpage}


\begin{thebibliography}{99}

\bibitem[Fregeau et al.(2004)]{fregeau04} Fregeau, J.~M., Cheung, P., Portegies Zwart, S.~F., Rasio, F.~A. 2004, \mnras, 352, 1

\bibitem[Geller \& Leigh(2015)]{geller15} Geller, A.~M., Leigh, N.~W.~C. 2015, ApJL, 808, 25

\bibitem[Ginat \& Perets(2021)]{GinatPerets2021} Ginat, Y.~B. \& Perets, H.~B.\ 2021, Physical Review X, 11, 031020. 

\bibitem[Harrington(1974)]{harrington74} Harrington, R.~S. 1974, Celestial Mechanics, 9, 465

\bibitem[Heggie(1975)]{heggie75} Heggie D.~C., 1975, MNRAS, 173, 729. doi:10.1093/mnras/173.3.729

\bibitem[Hoogerwerf et al.(2001)]{hoogerwerf01} Hoogerwerf, R., de Bruijne, J. H. J., de Zeeuw, P. T. 2001, A\&A, 365, 49

\bibitem[Leigh \& Sills(2011)]{leigh11} Leigh, N.~W.~C., Sills, A. 2011, \mnras, 410, 2370

\bibitem[Leigh \& Geller(2012)]{leigh12} Leigh, N.~W.~C., Geller, A.~M. 2012, \mnras, 425, 2369

\bibitem[Leigh \& Geller(2013)]{leigh13} Leigh, N.~W.~C., Geller, A.~M. 2013, \mnras, 432, 2474

\bibitem[Leigh et al.(2016)]{Leigh+16} Leigh, N.~W.~C., Stone, N.~C., Geller, A.~M., et al.\ 2016, \mnras, 463, 3311 

\bibitem[Mikkola(1983)]{mikkola83} Mikkola, S. 1983, \mnras, 203, 1107

\bibitem[Mikkola(1984a)]{mikkola84a} Mikkola, S. 1984, \mnras, 207, 115

\bibitem[Mikkola(1984b)]{mikkola84b} Mikkola, S. 1984, \mnras, 208, 75

\bibitem[Monaghan(1976a)]{Monaghan76a} Monaghan, J.~J.\ 1976a, \mnras, 176, 63 

\bibitem[Monaghan(1976b)]{Monaghan76b} Monaghan, J.~J.\ 1976b, \mnras, 177, 583 

\bibitem[Nash \& Monaghan(1980)]{nash80} Nash, P.~E., Monaghan, J.~J.\ 1980, \mnras, 192, 809

\bibitem[Perets \& Fabrycky(2009)]{perets09} Perets, H.~B., Fabrycky, D.~C. 2009, ApJ, 697, 1048 

\bibitem[Oh et al.(2015)]{oh15} Oh, A., Pavel, K., Pflamm-Altenburg, J. 2015, ApJ, 805, 92

\bibitem[Rasio, McMillan \& Hut(1995)]{rasio95} Rasio, F.~A., McMillan, S., Hut, P. 1995, \apjl, 438, L33

\bibitem[Ryu, Leigh \& Perna(2017a)]{ryu17a} Ryu, T., Leigh, N.~W.~C., Perna, R. 2017, \mnras, 467, 4447

\bibitem[Ryu, Leigh \& Perna(2017b)]{ryu17b} Ryu, T., Leigh, N.~W.~C., Perna, R. 2017, \mnras, submitted (arXiv:1703.08551)

\bibitem[Ryu, Leigh \& Perna(2017c)]{ryu17c} Ryu, T., Leigh, N.~W.~C., Perna, R. 2017, \mnras, submitted (arXiv:1703:08538)

\bibitem[Saslaw, Valtonen \& Aarseth(1974)]{saslaw74} Saslaw, W.~C., Valtonen, M.~J., Aarseth, S.~J. 1974, \apj, 196, 253

\bibitem[Sigurdsson \& Phinney(1993)]{sigurdsson93} Sigurdsson, S., Phinney, E.~S. 1993, ApJ, 415, 631

\bibitem[Stone \& Leigh(2019)]{stoneleigh19} Stone N.~C., Leigh N.~W.~C., 2019, Nature, 576, 406. doi:10.1038/s41586-019-1833-8

\bibitem[Valtonen \& Karttunen(2006)]{Valtonen&Karttunen06} Valtonen, M., \& Karttunen, H.\ 2006, The Three-Body Problem, by Mauri Valtonen and Hannu Karttunen, pp.~.~ISBN 0521852242.~Cambridge, UK: Cambridge University Press,  2006

\end{thebibliography}
\end{document}